\documentclass[acmsmall,nonacm]{acmart}

\setcopyright{acmlicensed}
\copyrightyear{2024}
\acmYear{2024}
\acmDOI{XXXXXXX.XXXXXXX}

\acmConference[Conference acronym 'XX]{Make sure to enter the correct
  conference title from your rights confirmation email}{June 03--05,
  2018}{Woodstock, NY}
\acmISBN{978-1-4503-XXXX-X/18/06}

\usepackage{listings}

\lstdefinelanguage{myasm}{
  morecomment=[l]{\#},
  morecomment=[l]{;},
  morecomment=[s]{/*}{*/},
  morestring=[b]",
  sensitive=false,
  morekeywords={movq,leaq,syscall,lock,cmpxchgq,vmovq,vaddsd,jmp,endbr64,vxorpd,vcomisd,ja,vaddsd,ret,vmovdqu16,vpbroadcastw,vpsubw,vporq,vzeroupper,movl,vmovss,fldt,fmulp,fstpt,xorq,addq,vmulsd,vmovsd,vmovupd,vmulpd,call,subq,jne,andpd,xorpd,movapd,cmplted,movsd,andpd,andnpd,orpd,addsd,retq,shrq,andl,subl,movabsq,andq,sall,vmovd,addl,orq,cmpltsd,xorl,rolq,xchgq,ucomisd,jbe,subsd,cmpxchg8b,popl,pushl,fldl,fadds,orl,fstpl,fildq,fistpq},
  alsoletter={.:\%},
  morekeywords={.quad,.text,.data,.long,.section,.value,.globl,.string,.p2align,.type},
  morekeywords=[2]{\%rip,\%rax,\%rdi,\%rsp,\%st,\%xmm0,\%xmm1,\%xmm2,\%xmm3,\%zmm0,\%rdx,\%eax,\%zmm1,\%rcx,\%edi,\%edx,\%rbx,\%rsi,\%ecx,\%esp,\%esi,\%ebx},
}
\lstdefinelanguage{vex}{
    keywords = {}
}
\lstdefinelanguage{ieee754}{
    keywords = {}
}
\lstdefinelanguage{asm}{ 
    keywords = {}
}

\usepackage{tikz}
\usepackage{pgfplots}
\pgfplotsset{compat=1.18} 
\usetikzlibrary{matrix,fadings,calc,positioning,decorations.pathreplacing,decorations.text,arrows,patterns,plotmarks,babel,mindmap,trees,shadows,tikzmark}
\usepgfplotslibrary{fillbetween}
\usetikzlibrary{external}
\newcommand{\tikzextchoice}[2]{#1} 
\tikzextchoice{}{\tikzexternalize}

\usepackage{siunitx}
\usepackage{multirow}
\usepackage{xcolor}

\floatstyle{plaintop}
\newfloat{lstfloat}{tbp}{lop}
\floatname{lstfloat}{Listing}

\newenvironment{customlegend}[1][]{%
    \begingroup
    \csname pgfplots@init@cleared@structures\endcsname
    \pgfplotsset{#1}%
}{%
    \csname pgfplots@createlegend\endcsname
    \endgroup
}%
\def\addlegendimage{\csname pgfplots@addlegendimage\endcsname}

\begin{document}

\newenvironment{codeq}[1]{
\def\codeqLabel{#1}
\par\vspace{\abovedisplayskip}
\begin{minipage}[c]{0.8\linewidth}
}{
\end{minipage}\hfill
\begin{minipage}[c]{0.09\linewidth}
   \flushright \refstepcounter{equation}\label{\codeqLabel}(\theequation)
\end{minipage}
\par\vspace{\belowdisplayskip}
}

\newcommand{\Cpp}{C\nolinebreak\hspace{-.05em}\raisebox{.4ex}{\tiny\bfseries +}\nolinebreak\hspace{-.10em}\raisebox{.4ex}{\tiny\bfseries +}}

\newcommand{\RR}{{\mathbb R}}
\newcommand{\EE}{{\mathbb E}}
\newcommand{\PP}{{\mathbb P}}
\newcommand{\dd}{{\mathrm d}}
\newcommand{\Var}{{\text{Var}}}
\newcommand{\OO}{{\mathcal O}}
\newcommand{\dotequal}{:=}
\newcommand{\expected}[1]{\EE #1}
\newcommand{\samplemean}[1]{\overline{#1}}
\newcommand{\expectedvaluediff}{\expected{(f')}}
\newcommand{\averagediff}{\samplemean{f'}}
\newcommand{\diffexpectedvalue}{(\expected{f})'}

\title{Efficient Forward-Mode Algorithmic Derivatives of Geant4}

\author{Max Aehle}
\email{max.aehle@scicomp.uni-kl.de}
\orcid{0000-0002-6739-5890}
\affiliation{%
  \institution{University of Kaiserslautern-Landau (RPTU)}
  \streetaddress{Gottlieb-Daimler-Straße}
  \city{Kaiserslautern}
  \country{Germany}
}

\author{Xuan Tung Nguyen}
\email{xuantung.nguyen@pd.infn.it}
\orcid{0009-0009-6527-441X}
\affiliation{%
  \institution{University of Kaiserslautern-Landau (RPTU)}
  \streetaddress{Gottlieb-Daimler-Straße}
  \city{Kaiserslautern}
  \country{Germany}
}
\affiliation{%
  \institution{National Institute for Nuclear Physics  (INFN)}
  \streetaddress{Marzolo street~8}
  \city{Padova}
  \country{Italy}
}

\author{Mihály Novák}
\email{mihaly.novak@cern.ch}
\affiliation{%
  \institution{European Organization for Nuclear Research (CERN)}
  \country{Switzerland/France}
}

\author{Tommaso Dorigo}
\email{tommaso.dorigo@gmail.com}
\orcid{0000-0002-1659-8727}
\affiliation{%
  \institution{Lulea University of Technology (LTU)}
  \streetaddress{Laboratorievagen 14}
  \city{Lulea}
  \country{Sweden}
}
\affiliation{%
  \institution{National Institute for Nuclear Physics  (INFN)}
  \streetaddress{Via Marzolo~8}
  \city{Padova}
  \country{Italy}
}


\author{Nicolas R.\ Gauger}
\email{nicolas.gauger@scicomp.uni-kl.de}
\orcid{0000-0002-5863-7384}
\affiliation{%
  \institution{University of Kaiserslautern-Landau (RPTU)}
  \city{Kaiserslautern}
  \country{Germany}
}

\author{Jan Kieseler}
\email{jan.kieseler@kit.edu}
\affiliation{%
  \institution{Karlsruhe Institute of Technology (KIT)}
  \city{Karlsruhe}
  \country{Germany}
}

\author{Markus Klute}
\email{markus.klute@kit.edu}
\affiliation{%
  \institution{Karlsruhe Institute of Technology (KIT)}
  \city{Karlsruhe}
  \country{Germany}
}

\author{Vassil Vassilev}
\email{vassil.vassilev@cern.ch}
\affiliation{%
  \institution{Princeton University}
  \city{Princeton}
  \state{New Jersey}
  \country{USA}
}

\renewcommand{\shortauthors}{Aehle, Nov\'ak, Vassilev, Gauger}

\setlength{\parindent}{0pt} 

\begin{abstract}

  We have applied an operator-overloading forward-mode algorithmic differentiation tool to the Monte-Carlo particle simulation toolkit Geant4. Our differentiated version of Geant4 allows computing mean pathwise derivatives of user-defined outputs of Geant4 applications with respect to user-defined inputs. This constitutes a major step towards enabling  gradient-based optimization techniques in high-energy physics, as well as other application domains of Geant4.

  This is a preliminary report on the technical aspects of applying operator-overloading AD to Geant4, as well as a first analysis of some results obtained by our differentiated Geant4 prototype. We plan to follow up with a more refined analysis.
\end{abstract}




\maketitle

\section{Introduction}

\subsubsection*{\bfseries Algorithmic Differentiation (AD)\@.} AD is a set of techniques to evaluate derivatives of computer-implemented functions, ideally in a mostly automatic fashion. To that end, AD tools can use a variety of mechanisms to identify all elementary real-arithmetic operations performed by the program to be differentiated, and combine their well-known derivatives according to the chain rule. 

For example, the \emph{forward mode} of AD with a single AD input variable $x$ keeps track of the \emph{dot value} $\dot a = \tfrac{\partial a}{\partial x}$ for every floating-point number $a$ appearing in the program, initialising $\dot x=1$, and $\dot c=0$ for every global variable or constant $c$, and augmenting every primal computation like $a_3 = a_1 \cdot a_2$ with a computation of the dot value of the result according to the appropriate differentiation rule, e.\,g.\ $\dot a_3 = \dot a_1 \cdot a_2 + a_1 \cdot \dot a_2$. See, e.\,g., Griewank and Walther \cite{griewank_evaluating_2008} for a general introduction to this field, including the \emph{reverse mode}. 

AD is essential for deep learning \cite{10.5555/3122009.3242010}, has been successfully used for aerodynamic engineering design \cite{doi:10.2514/6.2016-3518} and various other engineering design and parameter identification studies. Furthermore, AD is currently being investigated for use in the optimization of particle detectors, with applications envisioned in high-energy physics (HEP), astrophysics, nuclear and medical physics, and industrial use cases  \cite{NEURIPS2020_a878dbeb,DORIGO2023100085,10.1088/2632-2153/ad52e7,DORIGO2023167873}. 

\subsubsection*{\bfseries AD-Powered Detector Design.} On a high level, the idea behind end-to-end optimization of detectors is simple: Before a detector design is selected and realized, domain experts usually perform extensive studies based on computer simulations to assess various possible designs. Let us assume that designs can be described by a set of real-valued parameters $\theta\in\RR^n$, and that domain experts are able to express their preference for a given design by a scalar loss function value $f(\theta)\in\RR$, which can be evaluated by a (potentially very complex) computer program. While domain experts may check the value of $f(\theta)$ for a couple of
alternative designs $\theta$ 
and use their experience to interpolate in between, 
    such an approach is unlikely to find actual optima if the parameter space $\RR^n$ is high-dimensional. If the gradient $\nabla_\theta f(\theta)$ were known, it could be used by a gradient-based optimization algorithm to aid the experts in fine-tuning a good solution $\theta$ into an optimum. However, the code for $f$ may be a pipeline of complicated and partially stochastic computer programs, one of which is usually a simulation of the particle-detector interaction using the \emph{Geant4 toolkit for the simulation of the passage of particles through matter} \cite{agostinelli_geant4simulation_2003,allison_geant4_2006,allison_recent_2016}. While it is very desirable to extend this software with AD capabilities for detector optimization, applying AD in this context comes with technical and mathematical challenges.

\subsubsection*{\bfseries Technical Challenges.} AD can be applied to existing code, developed previously without AD in mind, by means of \emph{AD tools}. These tools can use different mechanisms to identify all real-arithmetic operations in the program and produce differentiated code, in a semi-automatic fashion. However, tools usually do not support the entire C{\ttfamily++} language standard and manual adaptations are therefore still required.

Applying AD to Geant4 is thus considered difficult, given Geant4's size and complexity, reflected in about one million lines of C{\ttfamily++} code. While we had already sucessfully applied the novel machine-code-based AD tool Derivgrind \cite{aehle_forward-mode_2022,aehle_reverse-mode_2022} to Geant4 with little manual efforts \cite{aehle2023progress}, the run-time of machine-code-based AD is currently about one order of magnitude worse than the run-time of classical implementations like the operator-overloading AD tool CoDiPack. 

\subsubsection*{\bfseries Mathematical Challenges.} On the mathematical side, it is well-known that a black-box application of an AD tool to a stochastic code, treating all random numbers like constants with regard to differentiation, does generally not lead to the types of derivatives required for optimization and other applications. 

\begin{lstfloat}
\caption{The function {\ttfamily f} samples from a Bernoulli distribution with success probability $c_1+c_2\theta$, using a stochastic primitive {\ttfamily flat()} that returns a random number uniformly distributed on $[0,1]$. Suppose that users want to minimize the expected value $\EE_\omega f(\theta,\omega)$, estimated by the Monte-Carlo iteration on the right. They are thus interested in $(\EE_\omega f(\theta,\omega))'=d$. However, applying AD to {\ttfamily Ef} yields an estimator for $\EE_\omega[f'(\theta,\omega)] = 0$.}
\label{lst:code-for-f}
\begin{minipage}{0.4\linewidth}
\begin{lstlisting}[language=c,basicstyle=\ttfamily,frame=single,mathescape]
double f(double theta){
  if(flat()<c1+c2*theta){
    return 1.0;
  } else {
    return 0.0;
  }
}
$~$
\end{lstlisting}
\end{minipage}\qquad
\begin{minipage}{0.4\linewidth}
\begin{lstlisting}[language=c,basicstyle=\ttfamily,frame=single]
double Ef(double theta){
  long const N = 1000;
  double sum = 0.0;
  for(long i=0; i<N; i++){
    sum += f(theta);
  }
  return sum/N;
}
\end{lstlisting}
\end{minipage}
\end{lstfloat}

Listing~\ref{lst:code-for-f} is a version a well-known counterexample \cite{arya2022automatic}: Given a parameter $\theta\in\RR$, {\ttfamily f} samples from the Bernoulli distribution with a success probability of $c_1 + c_2 \theta$, with some arbitrary constants $c_1 \in (0,1)$, $c_2\in\RR$. The Monte-Carlo algorithm {\ttfamily Ef} forms the average of many independent samples to estimate the expected value $\EE_\omega f(\theta,\omega)$. Note that in formulas, we indicate randomness by an additional argument $\omega\in \Omega$ from a random space $(\Omega, {\mathcal A}, \PP)$, while in the code, it is realized by a call {\ttfamily flat()} to a random number generator (RNG) that returns independent samples from the uniform distribution on $[0,1]$. The sample mean $\overline{\text{\ttfamily f}}$ computed by {\ttfamily Ef} approaches the expected value $\EE_\omega f(\theta,\omega) = c_1+c_2\theta$ for $\text{\ttfamily N}\to\infty$, with precise convergence statements available e.\,g.\ from the law of large numbers and the central limit theorem. Users seeking to minimize the expected value would be interested in its derivative, which we can analytically compute as
\begin{equation}\label{eq:diff-of-expected-value}
[\EE_\omega f(\theta,\omega)]' = [c_1+c_2\theta]' = c_2
\end{equation}
for this simple example. When AD is applied to {\ttfamily Ef}, treating random numbers as constants, the resulting differentiated function computes $\overline{\text{\ttfamily f}}' = \overline{\text{\ttfamily f}'}$, which approaches $\EE_\omega[f'(\theta,\omega)]$ for $\text{\ttfamily N}\to\infty$. Note that $f'(\theta,\omega)=0$ almost surely: $\theta$ does never directly affect the returned sample $f(\theta,\omega)$ in a differentiable fashion; it only affects the probabilities of the {\ttfamily if} and {\ttfamily else} branches being taken, which in turn affect the expected value of $f(\theta,\omega)$. Thus, using AD does not yield \eqref{eq:diff-of-expected-value} but 
\begin{equation}\label{eq:expected-value-of-diff}
\EE_\omega[f'(\theta,\omega)]= \EE_\omega[0] = 0.
\end{equation}

As a potential side problem, if the variance of $\nabla_\theta f(\theta, \omega)$ is too large, a computationally infeasible number of independent samples would be required to obtain a reliable average. And finally, it is known that the derivative of an approximative algorithm is not necessarily a good approximation for the derivative of the approximated function \cite{doi:10.1080/10556789208805503}. This being said, black-box AD may still provide good approximations of the derivatives required by the optimizer in practical applications, and small errors may be perfectly acceptable for the purpose of optimization and merely make the search through the parameter space slightly less efficient.

Aehle et al. \cite{ad-in-hepemshow} have recently tested this in a simple calorimetry setup similar to Geant4's \emph{TestEm3} example. Their study is concerned with mean pathwise derivatives of energy depositions in a sampling calorimeter with respect to parameters of the incoming particles and the detector geometry, computed by the HepEmShow \cite{hepemshow-github,hepemshow-doc} simulation based on the G4HepEm \cite{g4hepem-github} toolkit. Once a single physics process called \emph{multiple scattering} is disabled in the simulation, these mean pathwise derivatives have low variance, and approximate the derivative of the mean energy deposition with a small error in the order of \SI{5}{\percent}, which proved unproblematic in a simple optimization study.

\subsubsection*{\bfseries This Work.} As a quick study prior to the research presented in this paper, we have combined the works \cite{aehle2023progress} and \cite{ad-in-hepemshow}, which respectively targeted the technical and mathematical challenges: We used Derivgrind to compute the previously mentioned mean pathwise derivatives of energy depositions with respect to the primary energy and layer thicknesses, using a Geant4 simulation of the same detector geometry. As in \cite{ad-in-hepemshow}, the simulation of multiple scattering, i.\,e., particle deflections due to frequent electromagnetic interactions with atomic electric and magnetic fields, has been disabled. 
Encouraging results of this exploratory study then motivated us to integrate forward-mode operator-overloading AD into Geant4 to achieve higher performance. 

We report on the technical aspects of this integration (section~\ref{sec:ad-application}) and some results obtained for the TestEm3-like calorimeter geometry used by HepEmShow (section~\ref{sec:results}). This is a preliminary report; we yet have to integrate reverse-mode AD and explore the noisy behaviour of black-box derivatives of the Urban multiple scattering model and hadronic processes, among other possible improvements listed in \ref{sec:outlook}. {\bfseries So far, our differentiated version of Geant4 is a work-in-progress prototype and not well-tested.} Nevertheless, we invite Geant4 users who wish to explore electromagnetic interactions of particles with matter to test and use our Geant4 derivatives.

\section{Applying Operator-Overloading AD to Geant4}\label{sec:ad-application}

\subsection{Review of Operator-Overloading Algorithmic Differentiation}

Object-oriented programming languages allow to define new data types, and functions using these types. In particular, C{\ttfamily++} (and other languages) allow extending math operators like {\ttfamily +} and {\ttfamily *}, as well as function like {\ttfamily sqrt} and {\ttfamily exp}, to new datatypes. \emph{Operator-overloading AD tools} like CoDiPack \cite{SaAlGauTOMS2019} and ADOL-C \cite{Walther2012Gsw} use this language feature to provide AD types whose interface mimics built-in floating-point types like {\ttfamily double}, but which also track AD meta-data and perform AD logic alongside. Listing~\ref{lst:simple-ad-tool} sketches how an implementation of a simple operator-overloading AD type, and its {\ttfamily *} operator and {\ttfamily sqrt} function, might look like.

\begin{lstfloat}
\caption{Prototypical operator-overloading forward-mode AD tool, implemented as a C{\ttfamily++} header defining a forward-mode AD type {\ttfamily Forward} to be used instead of {\ttfamily double}.}
\label{lst:simple-ad-tool}
\begin{minipage}{0.9\linewidth}
\begin{lstlisting}[language=c++,basicstyle=\ttfamily,frame=single]
#include <math.h>

struct Forward {
  double value;
  double dot_value;
  Forward(double v): value(v), dot_value(0.0) {}
  Forward(): value(0.0), dot_value(0.0) {}
};

inline Forward operator*(Forward a, Forward b){
  Forward result{a.value * b.value};
  result.dot_value = a.dot_value * b.value
                     + a.value * b.dot_value;
  return result;
}

inline Forward sqrt(Forward a){
  Forward result{sqrt(a.value)};
  result.dot_value = 0.5/sqrt(a.value) * a.dot_value;
  return result;
}

/* ... define more operations and math functions ... */
\end{lstlisting}
\end{minipage}
\end{lstfloat}

Ideally, users wishing to apply an operator-overloading AD tool to their source code should only need to perform three steps:
\begin{itemize}
\item Exchanging the built-in floating-point datatype with the AD data type across the entire codebase, e.\,g.\ using the find-and-replace functionality of their code editor;
\item prepending source code files with {\ttfamily \#include} statements to the AD tool header file defining the AD type; and
\item indicating AD inputs and outputs in a tool-specific way.
\end{itemize}
In theory, the first two steps can be automated, and the effort required for the third step depends on the number of AD inputs and outputs but not on the overall code size.

In practice, however, applying operator-overloading AD to a large code base previously developed without AD in mind usually requires more work. This is partly because the C{\ttfamily++} standard makes subtle distinctions between built-in and class types, which prevent us from creating an AD type that could serve as a perfect drop-in substitute for {\ttfamily double}. We have set up a simple operator-overloading forward-mode AD tool \emph{EasyAD} to have precise control over the full interface of the AD type, and frequently adjusted it while fixing compiler errors after the type exchange in order to perform as few manual changes as possible in Geant4. In the next sections \ref{sec:considerations-easyad} and \ref{sec:considerations-geant4}, we outline code patterns in Geant4 that necessitated specific interface choices in the AD tool and manual adaptations in Geant4, respectively.

\subsection{Considerations Concerning the AD Type Interface}\label{sec:considerations-easyad}

In this section, we collect design considerations and requirements for an operator-overloading AD tool to allow for a simple integration with Geant4. Our implementation \emph{EasyAD} with the forward-mode AD type {\ttfamily Forward} is available at
\begin{center}
{\ttfamily https://github.com/SciCompKL/easyAD}
\end{center}

\subsubsection{\bfseries {\ttfamily constexpr} and literal types.}\label{sec:constexpr-literal} Geant4, and in particular its software dependency CLHEP, define {\ttfamily constexpr} {\ttfamily double} variables using simple arithmetic expressions on the right hand side, e.\,g.\ as in
\begin{equation}\label{eq:code-constexpr}
\text{\lstinline[language=c++,basicstyle=\ttfamily]|static constexpr G4double fTmin = 0.1 * CLHEP::keV;|}
\end{equation}
with $\text{\ttfamily G4double}=\text{\ttfamily double}$. To keep statements like~\eqref{eq:code-constexpr} well-formed with an AD type {\ttfamily Forward}, 
\begin{itemize}
    \item the AD type must be a \emph{literal type}, meaning that it has a trivial destructor and the appropriate {\ttfamily constexpr} constructor (with some more details provided in the standard), and
    \item the operations appearing on the right-hand side must be declared as {\ttfamily constexpr}.
\end{itemize}
While it is not difficult to follow these requirements in the forward mode, operations performed by a reverse-mode type may by default need access to a global tape, making it difficult to come up with a {\ttfamily constexpr} implementation. For future work, we suggest to either remove the {\ttfamily constexpr} specifier in statements like~\eqref{eq:code-constexpr} (and fix any errors caused by this), or keep the {\ttfamily double} type for these variables (meaning that they cannot serve as AD inputs).

\subsubsection{\bfseries AD Type for {\ttfamily float}.} Geant4 makes use of all of the three built-in C{\ttfamily++} floating-point types {\ttfamily float}, {\ttfamily double} and {\ttfamily long double}. Simply replacing them all by the same AD type would lead to compiler errors about function redefinitions, as previously different signatures like {\ttfamily void f(double)} and {\ttfamily void f(float)} would be replaced by the same signature {\ttfamily void f(Forward)}. We therefore provide a separate type {\ttfamily ForwardFloat} inheriting from {\ttfamily Forward} with a perfect forwarding constructor,
\begin{codeq}{code:ForwardFloat}
\begin{lstlisting}[language=c++,basicstyle=\ttfamily]
struct ForwardFloat : public Forward {
  template<typename...Args>
  ForwardFloat(Args&&...args): 
    Forward(std::forward<Args>(args)...) {}
};
\end{lstlisting}
\end{codeq}
Note that {\ttfamily ForwardFloat} now has the same floating-point accuracy as {\ttfamily Forward}, and likely a different accuracy than {\ttfamily float}. We do not expect that this affects the statistics of Geant4 outputs.

Additionally, we need to overload selected math functions for {\ttfamily float}s used by Geant4, like
\begin{codeq}{code:sqrtf}
\begin{lstlisting}[language=c++,basicstyle=\ttfamily]
inline ForwardFloat sqrtf(ForwardFloat a){
  return {sqrt(a.val), 0.5/sqrt(a.val) * a.dot};
}
\end{lstlisting}
\end{codeq}

Geant4 also rarely uses the {\ttfamily long double} type; we found that we can replace it with {\ttfamily Forward} instead of using a third AD type.

\subsubsection{\bfseries {\ttfamily std::complex}.} Geant4 occasionally uses the type {\ttfamily std::complex<double>}, which becomes {\ttfamily std::complex<Forward>} after the type exchange. The C{\ttfamily++} standards leave the behaviour of the latter unspecified (including the possibility of failing to compile) even if {\ttfamily Forward} is a \emph{NumericType}. Nevertheless, we did not encounter problems when using GCC and the GNU implementation of the C{\ttfamily++} standard library (libstdc{\ttfamily++}); merely, a few elementary operations needed to be explicitly overloaded, e.\,g.\  
\begin{codeq}{code:complex-mul}
\begin{lstlisting}[language=c++,basicstyle=\ttfamily]
inline std::complex<Forward> 
operator*(std::complex<Forward> a, double b){
  return a*Forward(b);
}
\end{lstlisting}
\end{codeq}
However, when compiling with the LLVM implementation of the C{\ttfamily++} standard library (libc{\ttfamily++}), {\ttfamily std::complex}-related errors occurred. 

\subsubsection{\bfseries Considerations regarding expression template.}\label{sec:expression-templates} In the operator-overloading AD tool CoDiPack \cite{SaAlGauTOMS2019}, the return type of math operations applied to AD types is not the AD type itself. Instead, operations return an expression type that stores information on the expression and its operands. This expression type is evaluated later, e.\,g.\ when passed as an input to assignment operators, or when explicitly converted to the AD type. This approach allows inserting AD logic on the level of entire expressions rather than single operations, e.\,g.\ to save tape space in the reverse mode. We do not use the expression template approach because it would make our AD tool and its interface significantly more complicated, and may easily lead to additional compiler errors after the type exchange. For example, with CoDiPack's forward-mode AD type, 
\begin{codeq}{code:ternary-codi}
\begin{lstlisting}[language=c++,basicstyle=\ttfamily]
codi::RealForward a = 1.0;
codi::RealForward b = (a > 0) ? (-a) : 0.0;
\end{lstlisting}
\end{codeq}
does not compile because the arguments  {\ttfamily (-a)} and {\ttfamily 0.0} of the ternary conditional operator have different types, none of which can be converted into the other. 

\subsubsection{\bfseries Casts from the AD type to {\ttfamily double}?} Geant4 sometimes assigns integer variables with floating-point values in order to perform a rounding operation. However, unlike {\ttfamily double}, our AD type does not provide implicit casts to integer types, for two reasons:
\begin{itemize}
    \item Constructs like
    \begin{codeq}{code:ternary-easyad}
    \begin{lstlisting}[language=c++,basicstyle=\ttfamily]
Forward a = 1.0;
Forward b = (a > 0) ? a : 0;
\end{lstlisting}
\end{codeq}
become ill-formed because the compiler cannot decide whether the return type of the ternary conditional operator should be {\ttfamily Forward} or an integer type.
    \item The compiler can combine casts like $\text{\ttfamily Forward}\to\text{\ttfamily int}$ with implicit conversions $\text{\ttfamily int} \to \text{\ttfamily double}$, making it easy to accidentally perform part of the real-arithmetic calculations without AD if some overloads are missing.
\end{itemize}
However, we provide explicit casts to any type that {\ttfamily double} can cast to:
\begin{codeq}{code:cast-explicit}
\begin{lstlisting}[language=c++,basicstyle=\ttfamily]
template<typename T>
explicit constexpr operator T() const {
  return val;
}
\end{lstlisting}
\end{codeq}

\subsubsection{\bfseries \boldmath What is $d \sqrt{x} / {dx}$ at $x=0$?} The square root function $\sqrt{x}$ is defined, but not differentiable, at $x=0$. Geant4 sometimes happens to evaluate $\sqrt{0}$, and we define the derivative to be zero there to avoid bringing NaNs into the computation.

\subsection{Changes in Geant4}\label{sec:considerations-geant4}

Our AD types {\ttfamily Forward} and {\ttfamily ForwardFloat} come close, but do not perfectly reproduce, the interface of {\ttfamily double} and {\ttfamily float}. Therefore, we performed additional manual modifications of the source code of Geant4 and some related codes following the type exchange to Geant4 version~11.0.4 described in paragraph~\ref{sec:type-exchange-in-geant4}. The differentiated codes are available on GitHub:
\begin{itemize}
\item the differentiated Geant4 version~11.0.4 source at {\ttfamily https://github.com/SciCompKL/geant4},
\item the differentiated CLHEP dependency at {\ttfamily https://github.com/SciCompKL/clhep},
\item the EasyAD-differentiated G4HepEm package at {\ttfamily https://github.com/SciCompKL/g4hepem} (branch {\ttfamily with\_easyad}).
\end{itemize}

Note that we arrived at these sets of modifications in a kind of debugging process, iteratively responding to compiler errors and later to unexpected derivative results by fixes in the above packages and in EasyAD. The remaining paragraphs give an overview of the changes we conducted. The non-automated changes amount to about one thousand lines of code. It may be worthwhile for future research to explore if some of these changes can be conducted automatically, following the ideas of H\"uck et al. \cite{HUCK20161485}.

\subsubsection{\bfseries Type Exchange}\label{sec:type-exchange-in-geant4} Large parts of the Geant4 source code already use type aliases $\text{\ttfamily G4double}=\text{\ttfamily double}$ and $\text{\ttfamily G4float}=\text{\ttfamily float}$, defined in {\ttfamily G4Types.hh}. We replace these alias definitions by 

\begin{codeq}{code:typeexchange-new}
\begin{lstlisting}[language=c++,basicstyle=\ttfamily]
#include "easyAD.hpp"
#include "easyAD_geant4extensions.hpp"
using G4double  = Forward;
using G4float = ForwardFloat;
\end{lstlisting}
\end{codeq}

Unfortunately, the type aliases are not consistently used across the entire code base of Geant4. We therefore used the following bash script to perform a type exchange:

\begin{codeq}{code:typeexchance-listing}
\begin{lstlisting}[language=bash,basicstyle=\ttfamily]
for suffix in c cc cpp icc h hh hpp; do
  find . -type f -not -path '*/\.*' -name "*.$suffix"  -exec sed -E -i -- 's/([^4])double/\1G4double/g' {} +
  find . -type f -not -path '*/\.*' -name "*.$suffix"  -exec sed -E -i -- 's/^double/G4double/g' {} +
  find . -type f -not -path '*/\.*' -name "*.$suffix"  -exec sed -E -i -- 's/([^4])float/\1G4float/g' {} +
  find . -type f -not -path '*/\.*' -name "*.$suffix"  -exec sed -E -i -- 's/^float/G4float/g' {} +
done
\end{lstlisting}
\end{codeq}
After this, we fixed a couple of wrong substitutions related to {\ttfamily std::ios::floatfield} and the headers {\ttfamily float.h} and {\ttfamily cfloat}. Also, we added missing type exchanges in a source file named {\ttfamily BooleanProcessor.src} and header files without suffix, e.\,g.\ in g4tools \cite{barrand2014softinex,hvrivnavcova2014integration}.

\subsubsection{\bfseries Removing Bit-Tricks}\label{sec:bit-tricks} As already reported by Aehle et al. \cite{aehle2023progress}, Geant4 implements a few math functions such as {\ttfamily G4Log} that operate on floating-point data in a binary fashion; these are generally not handled correctly by AD tools. We therefore replaced these functions by their standard math library counterparts.

\subsubsection{\bfseries Ambiguous overloads.} When the Geant4 function {\ttfamily G4UIparameter::SetDefaultValue} with multiple overloads
\begin{codeq}{code:ambiguous-setdefaultvalue}
\begin{lstlisting}[language=c++,basicstyle=\ttfamily]
void SetDefaultValue(const char* theDefaultValue);
void SetDefaultValue(G4int theDefaultValue);
void SetDefaultValue(G4long theDefaultValue);
void SetDefaultValue(G4double theDefaultValue);
\end{lstlisting}
\end{codeq}
is called with a {\ttfamily double} literal argument, the compiler reports an ambiguous overload because the argument could be converted both to integer types and to $\text{\ttfamily G4double} = \text{\ttfamily Forward}$. We fix this by explicitly casting the literal argument to {\ttfamily G4double}. Alternatively, one could add another overload for {\ttfamily double}.

\subsubsection{\bfseries Broken template argument type deduction.} Geant4 defines function templates such as 
\begin{codeq}{code:Divide}
\begin{lstlisting}[language=c++,basicstyle=\ttfamily]
template< class T >
void Divide( G4int Elements, T* To, T Numerator, 
  T* Denominator = NULL ){
  /* ... */
}
\end{lstlisting}
\end{codeq}
Such templates require that the floating-points types in the three arguments are the same; if a call is made using a {\ttfamily double} literal for {\ttfamily Numerator} but $\text{\ttfamily G4double}=\text{\ttfamily Forward}$ arrays for {\ttfamily To} and {\ttfamily Denominator}, it leads to a \emph{no matching function for call} compiler error.
Similarly, in calls to the standard library template function 
\begin{codeq}{code:accumulate}
\begin{lstlisting}[language=c++,basicstyle=\ttfamily]
template< class InputIt, class T >
T accumulate( InputIt first, InputIt last, T init );
\end{lstlisting}
\end{codeq}
a manual cast must be added to {\ttfamily double} literals passed for {\ttfamily init}.

\subsubsection{\bfseries Missing casts to integer types.} As outlined in the previous section~\ref{sec:considerations-easyad}, {\ttfamily Forward} does not define implicit casts to integers, only explicit casts defined by \eqref{code:cast-explicit}. Therefore, such explicit casts must be added in the Geant4 source code wherever indicated by compiler errors.

\subsubsection{\bfseries Unions.} We decided to have a non-trivial default constructor for {\ttfamily Forward}, which means that when this type is used in a {\ttfamily union}, the {\ttfamily union}'s default constructor is implicitly deleted. We therefore moved the {\ttfamily G4double} and {\ttfamily G4float} members of the union {\ttfamily tools::value::u} outside, and turned them into members of a new struct {\ttfamily tools::value::v}. It might be possible to explicitly define the constructor of {\ttfamily tools::value::u}.

\subsubsection{\bfseries Construction of Random Numbers.} CLHEP's implementation of the MIXMAX pseudo random-number generator \cite{SAVVIDY2015161} uses a function
\begin{codeq}{code:convert1double}
\begin{lstlisting}[language=c,basicstyle=\ttfamily]
inline double convert1double(myuint_t u) {
  const double one = 1;
  const myuint_t onemask = *(myuint_t*)&one;
  myuint_t tmp = (u>>9) | onemask; 
  double d = *(double*)&tmp;
  return d-1.0;
}
\end{lstlisting}
\end{codeq}
that initializes the sign and exponent bits of a {\ttfamily double d} with those of {\ttfamily 1.0}. The mantissa bits are copied from the argument {\ttfamily u}, which is a 64-bit integer with three leading zero bits and 61~random bits. This makes {\ttfamily d}  a random number uniformly distributed on the interval $[1,2]$.

After a simple type exchange, the program continues to initialize the value part of the {\ttfamily Forward d} with a random number, but the dot value of \lstinline+*(Forward*)&tmp+ is not set anywhere (and will likely match the data on the stack immediately behind \lstinline+&tmp+). We slightly modified this piece of code to make sure that the dot values of random numbers are properly set to zero.

\section{Results}\label{sec:results}

\begin{table}
    \centering
    \caption{Parameters of the sampling calorimeter geometry, image courtesy of Nov\'ak et al.\ \cite{hepemshow-github,hepemshow-doc}. }
    \label{tab:geometry}
    \begin{minipage}[c]{0.55\linewidth}
    \begin{tabular}{lccr}
         \toprule
         Parameter & Default value  \\
         
         \midrule
         Kinetic energy of primaries & \SI{10000}{\mega\eV} \\
         Thickness of absorber layers &  \SI{2.3}{\milli\meter} \\
         Thickness of gap layers &  \SI{5.7}{\milli\meter} \\
         Transversal dimension &  \SI{400}{\milli\meter} \\
         Number of layers & 50 \\
         Type of primary particles & electrons \\
         Absorber material & $\text{PbWO}_4$\\
         Gap material  & liquid Ar\\
         External mag.\ field & $(\SI{0}{\tesla},\SI{0}{\tesla},\SI{0}{\tesla})$ \\
         \bottomrule
    \end{tabular}
    \end{minipage}
    \begin{minipage}[c]{0.35\linewidth}
    \includegraphics[width=\textwidth]{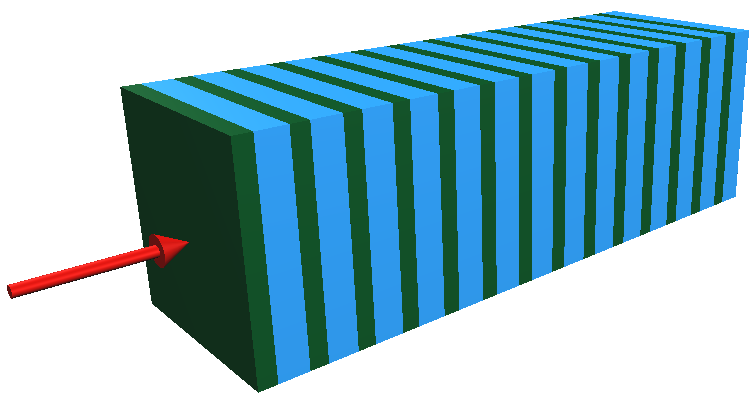} 
    \end{minipage}
\end{table}

\subsection{Reproducing the HepEmShow/G4HepEm findings with G4HepEm's TestEm3}\label{sec:results-g4hepem-testem}
Along with Geant4 and its CLHEP dependency, we have differentiated the G4HepEm toolkit \cite{g4hepem-github} and its example application called \emph{TestEm3}. G4HepEm's TestEm3 application has been adapted from, but is distinct to, the Geant4 example application of the same name differentiated below in section~\ref{sec:results-geant4-testem}. TestEm3 simulates showers in a sampling calorimeter with a simple parametric geometry displayed in Tab.~\ref{tab:geometry}. Note that the original absorber material $\text{Pb}$ has been replaced by $\text{PbWO}_4$, to cover the technical case with a material composed from multiple different chemical elements, which has a slightly larger radiation length and is commonly used in calorimeters (albeit as active material). The setup matches the one used in the previous study about AD in HepEmShow/G4HepEm \cite{ad-in-hepemshow}, which was mainly concerned with the derivatives of the energy depositions in the 50~layers with respect to the initial kinetic energy~$e$ of the incoming electrons, and the geometric thicknesses~$a$ and~$g$ of the absorber and gap layers, respectively. 

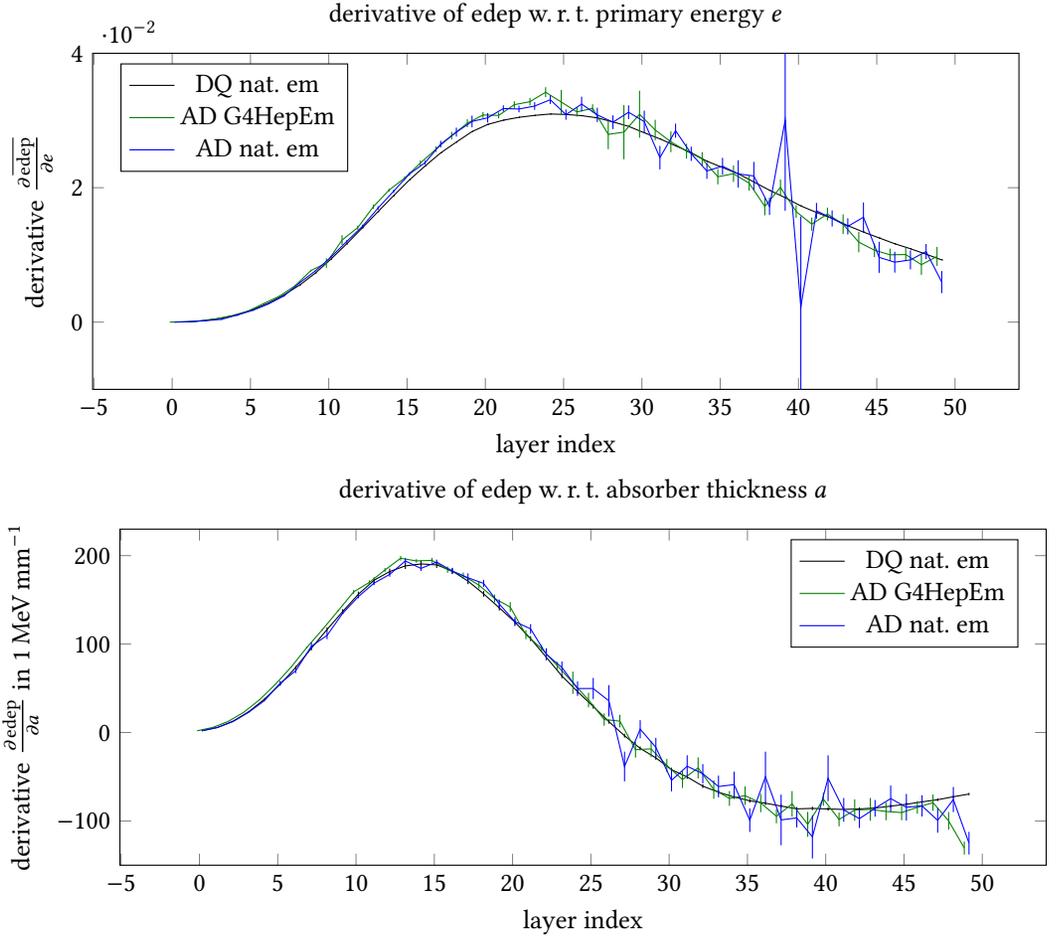
\begin{figure}
    \centering
    \tikzextchoice{\includegraphics{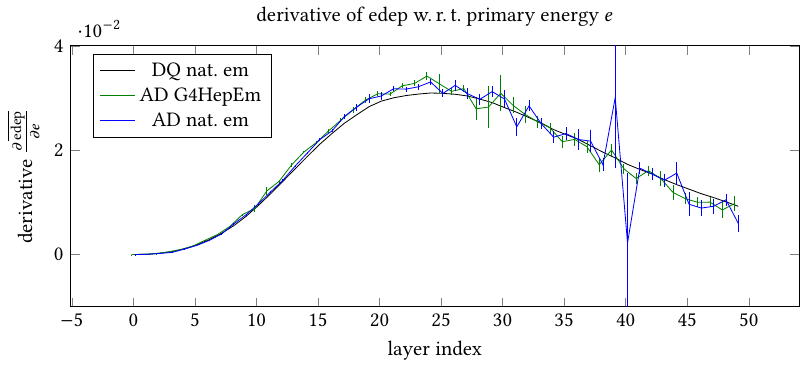}}{
    \tikzsetnextfilename{d_edep_d_primaryenergy}
    \begin{tikzpicture}
\begin{axis}[title={derivative of edep w.\,r.\,t.\ primary energy~$e$},ylabel={derivative $\tfrac{\partial\,\overline{\text{edep}}}{\partial e}$},ylabel style={align=center},ylabel near ticks,xlabel={layer index},height=6cm,error bars/y dir=both,error bars/y explicit,error bars/error bar style={thin},error bars/error mark=none,width=\linewidth,ymin=-0.01,ymax=0.04,legend pos=north west]
\addplot[black] table[x expr={\thisrowno{0}+0.15}, y index=1,y error index=2] {images/mean_std_de_g4em_num.dat};
\addplot[green!50!black] table[x expr={\thisrowno{0}-0.15}, y index=1,y error index=2] {images/mean_std_de_hepem.dat};
\addplot[blue] table[x expr={\thisrowno{0}+0.15}, y index=1,y error index=2] {images/mean_std_de_g4em.dat};
\addlegendentry{DQ nat.\ em}
\addlegendentry{AD G4HepEm}
\addlegendentry{AD nat.\ em}
\end{axis}
\end{tikzpicture}
}
\tikzextchoice{\includegraphics{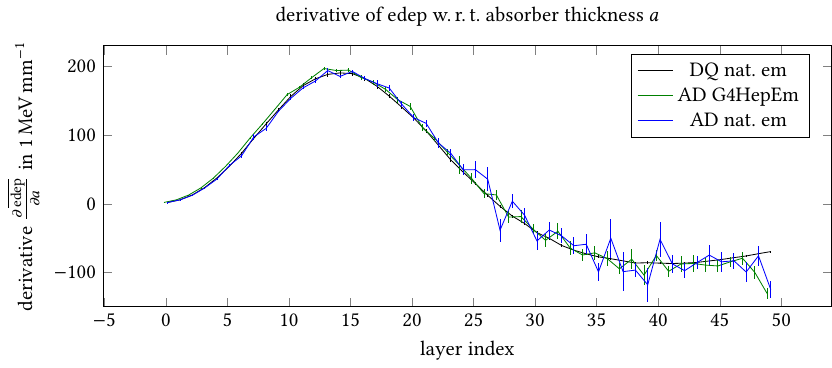}}{
    \tikzsetnextfilename{d_edep_d_absorberthickness}
    \begin{tikzpicture}
\begin{axis}[title={derivative of edep w.\,r.\,t.\ absorber thickness~$a$},ylabel={derivative $\tfrac{\partial\,\overline{\text{edep}}}{\partial a}$ in \SI{1}{\mega\eV\per\milli\meter}},ylabel style={align=center},ylabel near ticks,xlabel={layer index},height=6cm,error bars/y dir=both,error bars/y explicit,error bars/error bar style={thin},error bars/error mark=none,width=\linewidth,ymin=-150,ymax=230,legend pos=north east]
\addplot[black] table[x expr={\thisrowno{0}+0.15}, y index=1,y error index=2] {images/mean_std_da_g4em_num.dat};
\addplot[green!50!black] table[x expr={\thisrowno{0}-0.15}, y index=1,y error index=2] {images/mean_std_da_hepem.dat};
\addplot[blue] table[x expr={\thisrowno{0}+0.15}, y index=1,y error index=2] {images/mean_std_da_g4em.dat};
\addlegendentry{DQ nat.\ em}
\addlegendentry{AD G4HepEm}
\addlegendentry{AD nat.\ em}
\end{axis}
\end{tikzpicture}
}
    \caption{Algorithmic derivatives of the edep with respect to the primary energy~$e$ and the absorber thickness~$a$ for electromagnetic showers, computed using the differentiated TestEm3 Geant4 applications from the G4HepEm repository, using either G4HepEm or the native Geant4 electromagnetic physics. Difference quotients with G4HepEm physics are very close to those with the native physics.}
    \label{fig:d-edep-d-stuff-g4}
\end{figure}

In Fig.~\ref{fig:d-edep-d-stuff-g4}, computed using 2.4\,M events, we reproduce the observation of \cite{ad-in-hepemshow} that the pathwise gradient estimator has a low variance and bias once multiple scattering has been disabled in the simulation.

Specifically, the green lines in Fig.~\ref{fig:d-edep-d-stuff-g4} show the derivatives of a Geant4 simulation using G4HepEm to sample electromagnetic physics. Compared to the previous study \cite{ad-in-hepemshow} that relied on a Geant4-independent geometry definition and stepping loop from the HepEmShow package, the present setup shows that the corresponding implementations in Geant4, which follow the same algorithmic ideas, again do not pose problems to AD.

For the blue lines in Fig.~\ref{fig:d-edep-d-stuff-g4}, the G4HepEm physics have been replaced by a list of the respective native Geant4 physics:
\begin{codeq}{code:g4em-physics}
\begin{center}
\begin{minipage}{0.95\textwidth}
{\ttfamily G4PhotoElectricEffect},
{\ttfamily G4ComptonScattering},
{\ttfamily G4GammaConversion},
{\ttfamily G4eIonisation},
{\ttfamily G4eBremsstrahlung},
{\ttfamily G4eplusAnnihilation}
\end{minipage}
\end{center}
\end{codeq}
As in \cite{ad-in-hepemshow}, multiple scattering has been disabled. Note that G4HepEm only covers interactions of electrons, positrons and photons, and therefore the list~\eqref{code:g4em-physics} misses gamma-nuclear and lepto-nuclear processes that may produce other types of particles. As the plots show, replacing G4HepEm with the respective native Geant4 processes does not introduce problems with respect to AD.

\subsection{Exploring Wider Ranges of Physics Processes with Geant4's TestEm3}\label{sec:results-geant4-testem}
We have also differentiated the original Geant4's TestEm3 application. Explicitly setting the {\ttfamily "FTFP\_BERT"} physics list in the code (which also includes hadronic interactions), we can control the set of simulated physics processes by editing the Geant4 source files
\begin{codeq}{code:ftfp-paths}
\begin{lstlisting}[mathescape,breaklines=true]
source/physics_lists/lists/src/FTFP_BERT.cc $\text{\quad \emph{and}}$
source/physics_lists/constructors/electromagnetic/src/G4EmStandardPhysics.cc $.$
\end{lstlisting}
\end{codeq}
Starting with the full {\ttfamily FTFP\_BERT} physics list, we obtain noisy derivatives.

\subsubsection{Disabling Multiple Scattering} When the MSC-related models 
\begin{codeq}{code:msc-disabled}
\begin{center}
\begin{minipage}{0.8\textwidth}
\lstinline|G4hMultipleScattering("ionmsc")|, \lstinline|G4UrbanMscModel|, \\
\lstinline|G4WentzelVIModel|, \lstinline|G4CoulombScattering|
\end{minipage}
\end{center}
\end{codeq}
are disabled in {\ttfamily G4EmStandardPhysics.cc}, we usually obtain plausible algorithmic derivatives (of the energy depositions with respect to the primary energy) similar to those in Fig.~\ref{fig:d-edep-d-stuff-g4} when the number of events is moderate (e.\,g.\ 1000). Thus, all standard electromagnetic processes except those listed above seem to be out-of-the-box compatible with black-box AD.

\subsubsection{Alternative to disabling MSC\@.}\label{sec:msc-alternatives} Analyzing the noisy derivatives in the G4HepEm/HepEmShow setup \cite{ad-in-hepemshow} in depth, we found that large-magnitude (noisy) derivatives mainly originate from the deflection angle, whose cosine is sampled by 
\begin{codeq}{code:cost-files}
\begin{center}
\begin{minipage}{0.999\textwidth}
{\ttfamily G4HepEmElectronInteractionUMSC::SampleCosineTheta} in G4HepEm, and \\ 
{\ttfamily G4UrbanMscModel::SampleCosineTheta} in Geant4.
\end{minipage}
\end{center}
\end{codeq}
When we explicitly set the dot value of this number to zero, we can re-enable all the multiple scattering models disabled in \eqref{code:msc-disabled} and still obtain non-noisy derivatives with Geant4's TestEm3 application (though, qualitatively, with a somewhat higher bias). This way it is possible to include MSC in the simulation, but neglect some dependencies created by MSC for the derivative computation. 

\subsubsection{Hadronic primaries} When the incoming primary particles are protons, neutrons, $\pi^+$, $\pi^-$, $K^+$, $K^-$, $K^0$ or $\bar K^0$ instead of electrons, positrons and gammas, derivatives are very noisy, exceeding the plausible range for the derivatives by tens of orders of magnitude. 

\subsubsection{Non-standard electromagnetic processes} Back to a version with MSC processes in \eqref{code:msc-disabled} disabled, we also noticed that with a large number of simulated events (e.\,g.\ 100\,k), the derivatives usually become noisy as well. We attribute this to rare gamma-nuclear and lepto-nuclear processes that produce types of particles other than electrons, positrons and gammas, and the subsequent interactions of those particles. When we disable the {\ttfamily G4EmExtraPhysics}-list in {\ttfamily FTFP\_BERT.cc}, we again obtain plausible and non-noisy derivative values (even for 1\,M events).

\section{Outlook}\label{sec:outlook}

\subsection{Technical Improvements and Extensions}
An important next step on the technical side is to provide reverse-mode AD, as it would make the run-time required to obtain an entire gradient $\nabla_\theta \overline{f(\theta)} \in \RR^{n\times 1}$ independent from the number $n$ of design parameters. In principle, reverse-mode AD can be implemented in a similar fashion as operator-overloading forward-mode AD, replacing the dot-value-propagating logic by code that handles indices and records the real-arithmetic evaluation tree on a data structure called the \emph{tape}. It is, however, not straightforward to implement a reverse-mode AD type as a C{\ttfamily++} literal type (paragraph~\ref{sec:constexpr-literal}), so further adaptations in Geant4 might be necessary.  Additionally, as the tape can quickly grow very large, it is advisable to employ advanced AD workflows in Geant4, like evaluating and clearing the tape after each Monte-Carlo iteration rather than at the end \cite{Hascoet2002}.

For performance reasons and in order to leverage a wider set of AD capabilities, it would be nice to switch to a more mature operator-overloading AD tool like CoDiPack \cite{SaAlGauTOMS2019}. However, as the {\ttfamily Forward} type has been developed with the main goal to keep its integration easy, tools focusing on high performance might require more severe modifications of Geant4's source code (paragraph~\ref{sec:expression-templates}). To that end, tools that automatically replace code constructs incompatible with operator overloading could be explored \cite{HUCK20161485}. 

Likewise, it would be interesting to see source-transformation AD tools like Clad \cite{Vassilev_Clad} being applied to Geant4, as they can access more information on the program to be differentiated and may therefore offer higher performance and smaller tape sizes.

\subsection{Further Insight in Physics Details}

Like Aehle et al.\ \cite{ad-in-hepemshow}, we disabled multiple scattering in the simulation as the pathwise derivatives were very noisy otherwise. We presented a fix in section~\ref{sec:msc-alternatives}, but it would be good to reach a more detailed understanding. 

Similarly, we currently lack the capability to estimate gradients in a meaningful manner when simulating hadrons. As pointed out by Aehle et al.\ \cite{ad-in-hepemshow}, this does not imply that the underlying physical processes are intrinsically badly-suited for gradient-based optimization, but certainly the respective models and algorithms need to be revisited in order to try to make them more AD-friendly. 

\subsection{Testing and Application}
While we have only worked with variants of the TestEm3-like HepEmShow testcase so far, forward-mode AD has been applied to the entire code base of Geant4 version~11.0.4. It should therefore be equally possible for a wide variety of Geant4-based applications to
\begin{itemize}
    \item differentiate the application itself in a similar way as the TestEm3 example application described above in sections~\ref{sec:results-g4hepem-testem}, \ref{sec:results-geant4-testem},
    \item identify simulation parameters and outputs, such that the derivative of the outputs with respect to the parameters can be useful, and
    \item compare these algorithmic derivatives against difference quotients.
\end{itemize}
Even though our differentiated version of Geant4 is still work in progress, we encourage everyone interested to study Geant4 derivatives for their setup of interest:

\begin{center}
    \ttfamily https://github.com/SciCompKL/geant4
\end{center}

\section{Acknowledgments}
The authors would like to thank Alberto Ribon for valuable discussions on differentiability in high-energy physics.

MA and NG gratefully acknowledge the funding of the research training group SIVERT by the German federal state of Rhineland-Palatinate. Also, MA and NG gratefully acknowledge the funding of the German National High Performance Computing (NHR) association for the Center NHR South-West. JK is supported by the Alexander-von-Humboldt-Stiftung. VV is supported by the National Science Foundation under Grant OAC-2311471.

This research was supported by the Munich Institute for Astro-, Particle and BioPhysics (MIAPbP), which is funded by the Deutsche Forschungsgemeinschaft (DFG, German Research Foundation) under Germany's Excellence Strategy -- EXC-2094-390783311.
Computing resources have been provided by the Alliance for High Performance Computing in Rhineland-Palatinate (AHRP) via the Elwetritsch cluster at the University of Kaiserslautern-Landau.

\bibliographystyle{ACM-Reference-Format}
\bibliography{main}

\end{document}